\documentclass[lettersize,journal]{IEEEtran}
\usepackage{amsmath,amsfonts}
\usepackage{algorithmic}
\usepackage{array}
\usepackage[caption=false,font=normalsize,labelfont=sf,textfont=sf]{subfig}
\usepackage{textcomp}
\usepackage{stfloats}
\usepackage{url}
\usepackage{verbatim}
\usepackage{graphicx}
\def\BibTeX{{\rm B\kern-.05em{\sc i\kern-.025em b}\kern-.08em
    T\kern-.1667em\lower.7ex\hbox{E}\kern-.125emX}}
\usepackage{balance}

\usepackage{mathtools}
\usepackage{dsfont}
\DeclareMathOperator*{\argmax}{arg\,max}

\usepackage{xcolor}

\begin{document}

\title{Experimental Verification of PCH-EM Algorithm for Characterizing DSERN Image Sensors}
\author{Aaron Hendrickson, David P.~Haefner, Nicholas R.~Shade \IEEEmembership{(Student Member, IEEE)}, and Eric R.~Fossum \IEEEmembership{(Life Fellow, IEEE)}
\thanks{}}

\markboth{Journal of the Electron Devices Society,~Vol.~X, No.~X, Month~2023}%
{How to Use the IEEEtran \LaTeX \ Templates}

\maketitle

\begin{abstract}
The Photon Counting Histogram Expectation Maximization (PCH-EM) algorithm has recently been reported as a candidate method for the characterization of Deep Sub-Electron Read Noise (DSERN) image sensors. This work describes a comprehensive demonstration of the PCH-EM algorithm applied to a DSERN capable quanta image sensor. The results show that PCH-EM is able to characterize DSERN pixels for a large span of quanta exposure and read noise values. The per-pixel characterization results of the sensor are combined with the proposed Photon Counting Distribution (PCD) model to demonstrate the ability of PCH-EM to predict the ensemble distribution of the device. The agreement between experimental observations and model predictions demonstrates both the applicability of the PCD model in the DSERN regime as well as the ability of the PCH-EM algorithm to accurately estimate the underlying model parameters.
\end{abstract}




\section{Introduction}
\label{sec:introduction}

\IEEEPARstart{A}{s} the detection precision of advanced camera technology improves, the ability to properly characterize and evaluate modern image sensors only becomes more important. While the traditional Photon Transfer (PT) method \cite{beecken_96,janesick_2007,hendrickson_22} can be applied to Deep Sub-Electron Read Noise (DSERN) image sensors, it has been shown there are other algorithms that can improve the accuracy and precision of the camera characterization \cite{starkey_2016,Nakamoto_2022,hendrickson_2023}. Specifically, both the Photon Counting Histogram (PCH) method \cite{starkey_2016,fossum_2015_2,fossum_2015,Dutton_2016} and recently introduced Maximum Likelihood Estimation (MLE) based method \cite{Nakamoto_2022} have been demonstrated to incur less uncertainty in their estimates as compared to the PT method. Recently, Hendrickson and Haefner proposed a fourth method, Photon Counting Histogram Expectation Maximization (PCH-EM), that improves on these techniques by providing an automated algorithm for simultaneous maximum likelihood estimation of quanta exposure, conversion gain, bias (DC offset), and read noise of DSERN pixels from a single sample of data \cite{hendrickson_2023}.

Due to the cutting edge nature of DSERN capable sensors, the PCH-EM algorithm was initially demonstrated using simulated Monte Carlo experiments. In this paper, through the use of an early photon-counting-capable Quanta Image Sensor (QIS) from Gigajot Technology Inc., a more comprehensive demonstration of the PCH-EM algorithm and verification of the associated Photon Counting Distribution (PCD) model is provided. This is accomplished by first reviewing the assumed mathematical model and theoretical framework behind the PCH-EM method. New theory pertaining to ensemble statistics of DSERN sensors is also introduced. Experimental conditions and data capture methods needed for dark current characterization with PCH-EM are provided. The experimental observations are evaluated through the PCH-EM algorithm, providing a full characterization of the sensor giving per-pixel estimates of dark current, conversion gain, bias, and read noise all from a single sequence of images captured under dark conditions. The  per-pixel characterization results are then combined with the PCD model to predict the ensemble distribution for the device, showing that the model is able to predict the distribution of the raw sensor data. This agreement demonstrates both the applicability of the PCD model in the DSERN regime as well as the PCH-EM algorithm's ability to accurately estimate the underlying model parameters.


\section{Theory}
\label{sec:theory}


\subsection{The PCD}

The digital output of a DSERN pixel is modeled as
\begin{equation}
\begin{aligned}
    \label{eq:PCD_RV_definition}
    X &=\lceil (K+R)/g+\mu\rfloor\\
    K &\sim\operatorname{Poisson}(H)\\
    R &\sim\mathcal N(0,\sigma_R^2),
\end{aligned}
\end{equation}
where $H$ is the quanta exposure $(e\text{-})$, $\sigma_R$ the input referred analog read noise $(e\text{-})$, $g$ the conversion gain $(e\text{-}/\mathrm{DN})$, $\mu$ is the pixel bias $(\mathrm{DN})$, and $\lceil\cdot\rfloor$ denotes rounding to the nearest integer. As such, the random variable $X$ represents the random process of adding noise $(R)$ to a number of electrons $(K)$ followed by the application of gain, offset, and finally quantization. Note that this is a general sensor model not specific to DSERN devices. What differentiates DSERN pixels is the fact that the signal corrupting noise $R$ is sufficiently small so that the \emph{electron number} $K$ can be reasonably estimated.

Assuming $g\ll\sigma_R$, quantization (rounding) in (\ref{eq:PCD_RV_definition}) can be modeled as an additive noise component so that the distribution of $X$ is reasonably approximated by the \emph{Photon Counting Distribution} (PCD) \cite{hendrickson_2023}
\begin{equation}
    \label{eq:X_density}
    f_X(x|\theta)=\sum_{k=0}^\infty\frac{e^{-H}H^k}{k!}\phi(x;\mu+k/g,\sigma^2),
\end{equation}
where $\theta=(H,g,\mu,\sigma^2)$ are the PCD parameters and $\phi(x;\mu,\sigma^2)=\frac{1}{\sqrt{2\pi\sigma^2}}\exp(-(x-\mu)^2/2\sigma^2)$ is the Gaussian probability density with mean $\mu$ and variance $\sigma^2$. In (\ref{eq:X_density}), $\sigma=(\sigma_R^2/g^2+\sigma_Q^2)^{1/2}$ is the combined read and quantization noise in $(\mathrm{DN})$.


\subsection{The PCH-EM Algorithm}
\label{subsec:The_PCH-EM_algorithm}

Given a random sample $\mathbf x=\{x_1,\dots,x_N\}$ with $x_n\overset{\mathrm{iid}}{\sim}\operatorname{PCD}(H,g,\mu,\sigma^2)$ and an initial estimate of the parameters $\theta_0=(H_0,g_0,\mu_0,\sigma_0^2)$, the PCH-EM algorithm iteratively updates the parameter estimates via the update equations \cite{hendrickson_2023}
\begin{subequations}\label{eq:update_eqns}
\begin{align}
H_{t+1} &=A_t \label{eq:H_update}\\
g_{t+1}&=\frac{B_t-H_{t+1}^2}{C_t-\bar xH_{t+1}} \label{eq:g_update}\\
\mu_{t+1}&=\bar x-\frac{H_{t+1}}{g_{t+1}} \label{eq:mu_update}\\
\sigma_{t+1}^2 &= \overline{x^2}-\bar x^2-\frac{B_t-H_{t+1}^2}{g_{t+1}^2},\label{eq:sigma_update}
\end{align}
\end{subequations}
where $\bar x=\frac{1}{N}\sum_{n=1}^Nx_n$ and $\overline{x^2}=\frac{1}{N}\sum_{n=1}^Nx_n^2$ are the first two sample moments and
\begin{subequations}\label{eq:update_matrices}
\begin{align}
A_t &=\frac{1}{N}\sum_{n=1}^N\sum_{k=0}^\infty\gamma_{nk}^{(t)}k \label{eq:A_t}\\
B_t &=\frac{1}{N}\sum_{n=1}^N\sum_{k=0}^\infty\gamma_{nk}^{(t)}k^2 \label{eq:B_t}\\
C_t &=\frac{1}{N}\sum_{n=1}^Nx_n\sum_{k=0}^\infty\gamma_{nk}^{(t)}k, \label{eq:C_t}
\end{align}
\end{subequations}
where
\begin{equation}
    \label{eq:gamma_nk}
    \gamma_{nk}^{(t)}=\frac{\frac{e^{-H_t}H_t^k}{k!}\phi(x_n;\mu_t+k/g_t,\sigma_t^2)}{\sum_{\ell=0}^\infty\frac{e^{-H_t}H_t^\ell}{\ell!}\phi(x_n;\mu_t+\ell/g_t,\sigma_t^2)},
\end{equation}
are the so-called \emph{membership probabilities}; representing a probability distribution of the unknown electron number associated with each observation $x_n$. As such, the membership probabilities satisfy $\sum_{k=0}^\infty\gamma_{nk}^{(t)}=1$.

In each iteration, the algorithm takes the current estimate $\theta_t$ and then performs an Expectation (E) step to compute the $\gamma_{nk}^{(t)}$ followed by a Maximization (M) step to update the estimate according to (\ref{eq:update_eqns}). In doing so, the algorithm guarantees an increase in the likelihood of the sample at each iteration such that a local maxima of the likelihood function is always achieved\footnote{Assuming the starting point $\theta_0$ is sufficiently good, PCH-EM achieves the global maximum of the likelihood function so that the final estimates are maximum likelihood estimates for their respective parameters.} \cite{dempster_1977}. The algorithm halts when a specified convergence criteria is met.

In the context of machine learning, the general EM algorithm can be viewed as a density-based clustering algorithm, assigning labels to each datapoint based on what cluster the datapoint is most likely to belong to. In the context of PCH-EM, the Gaussian components comprising the PCD are the clusters, with the electron number determining which cluster an observation belongs. As such, a natural byproduct of the PCH-EM algorithm is the ability to map each observation $x_n$ to a nonnegative integer $\tilde k_n$ representing a best estimate for the electron number associated with each observation. This post-process \emph{denoising} of the sensor data is accomplished by applying the membership probabilities via
\begin{equation}
    \label{eq:demarg_K_eqn}
    \tilde k_n=\argmax_k\gamma_{nk}^{(t)}.
\end{equation}
In essence, the mapping $\tilde k_n:x_n\to\Bbb N_0$ described in (\ref{eq:demarg_K_eqn}) is clustering the data by its mostly likely electron number in an optimal manner as to reduce bit error rates \cite{fossum_2016_2}.

\begin{figure}[htb]
\centering
\includegraphics[scale=0.9]{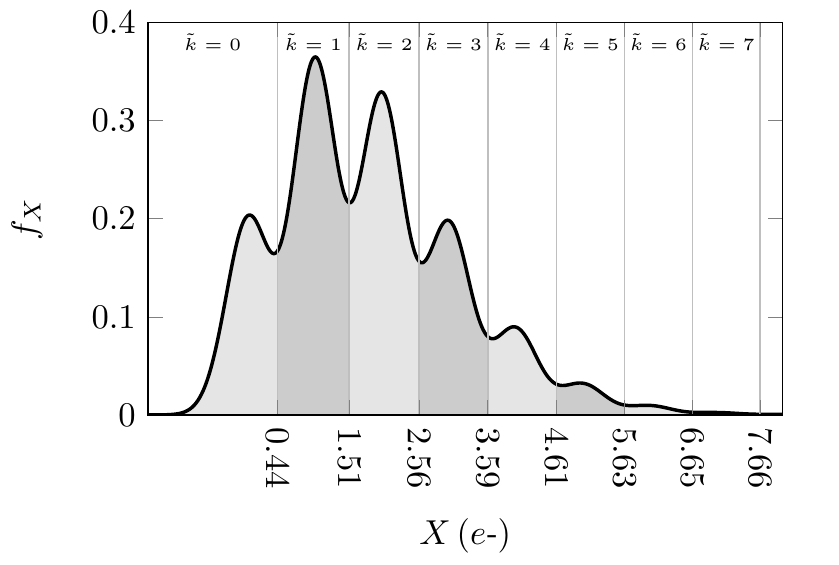}
\caption{Example PCD showing optimal cluster edges obtained through (\ref{eq:demarg_K_eqn}).}
\label{fig:optimal_quantization}
\end{figure}

To see this optimal clustering in action, consider the example $x_n\sim\operatorname{PCD}(1.8,1,0,(0.33)^2)$, where the values of the parameters $g=1$ and $\mu=0$ are selected so that the data can be interpreted as being in units of $e\text{-}$. Figure \ref{fig:optimal_quantization} shows the PCD along with the optimal cluster edges obtained from the quantization described in (\ref{eq:demarg_K_eqn}). While the PCD peaks are centered at nonnegative integers, it can be seen that the cluster edges are not directly centered between the peaks nor are the clusters of equal size. This nonuniform clustering ensures optimal estimation of the electron number for each observation.


\subsection{Ensemble Statistics}

The PCD in (\ref{eq:X_density}) describes the distribution of data produced by a single DSERN pixel. When considering data produced by an array of DSERN pixels, each with potentially different parameters, the parameters themselves can be modeled as random variables. Denoting $E$ as the random variable describing the ensemble of pixels leads to the hierarchical model
\begin{equation}
\begin{aligned}
    \label{eq:EPCD_RV_definition}
    E|H,g,\mu,\sigma^2 &\sim\operatorname{PCD}(H,g,\mu,\sigma^2)\\
    (H,g,\mu,\sigma^2) &\sim F_\theta
\end{aligned}
\end{equation}
so that the distribution of $E$ is given by the Ensemble PCD (EPCD)
 \begin{equation}
 \label{eq:full_ensemble_dist}
     f_E(x)=\iiiint_\Theta f_X(x|\theta)f_\theta(\theta)\,\mathrm d\theta,
 \end{equation}
where $F_\theta$ is the joint distribution of the parameters (with corresponding joint density $f_\theta$) and $\Theta\subset\Bbb R^4$ is the parameter space denoting all possible values of the parameter vector $\theta$. Alternatively, the EPCD can be written as $f_E(x)=\mathsf E_\theta(f_X(x|\theta))$, where $\mathsf E_\theta$ denotes the expected value w.r.t.~$\theta$. The moments of the EPCD can be given in terms of the moments of the parameters as shown in Appendix \ref{sec:ensemble_dist_moments}.

Unlike the per-pixel PCD, the peaks (local maxima) in the EPCD typically disappear at higher signal levels (c.f.~Figure \ref{fig:ensemble_plot} (top)), which is indicative of conversion gain nonuniformity (see Appendix \ref{sec:ensemble_peak_resolution}). For this reason, it is also useful to consider the ensemble distribution after correcting conversion gain nonuniformity and bias through a conventional two point Non-Uniformity Correction (NUC). Applying a gain and offset correction both improves the resolution of the peaks and centers them on the nonnegative integers. The Non-Uniformity Corrected (NUCed) EPCD can be found by setting $g=1$ and $\mu=0$ as constants leading to the model
\begin{equation}
\begin{aligned}
    \label{eq:NUCed_EPCD_RV_definition}
    E^\prime|H,\sigma_{e\text{-}}^2 &\sim\operatorname{PCD}(H,1,0,\sigma_{e\text{-}}^2)\\
    (H,\sigma_{e\text{-}}^2) &\sim F_{\theta^\prime}
\end{aligned}
\end{equation}
with distribution
\begin{equation}
\label{eq:partial_ensemble_dist}
    f_{E^\prime}(x)=\iint_{\Theta^\prime} f_X(x|H,1,0,\sigma_{e\text{-}}^2)f_{\theta^\prime}(\theta^\prime)\,d\theta^\prime,
\end{equation}
where $\sigma_{e\text{-}}=\sigma\times g$ is the total read and quantization noise in units of electrons and $\theta^\prime=(H,\sigma_{e\text{-}}^2)$. Examples of both the EPCD and NUCed EPCD can be seen in Figure \ref{fig:ensemble_plot} (see Section \ref{subsec:experimental_ensemble_dist}).

Lastly, consider the ensemble distribution of the electron number $K$, which will appear later when evaluating the ability of PCH-EM to predict electron numbers. On a per-pixel basis the electron number is Poisson distributed and since each pixel may have a unique quanta exposure, the ensemble electron number $K_e$ is described by
\begin{equation}
\begin{aligned}
    \label{eq:ensemble_K_definition}
    K_e|H &\sim\operatorname{Poisson}(H)\\
    H &\sim F_H
\end{aligned}
\end{equation}
with probability mass
\begin{equation}
    \label{eq:K_ensemble_pmf}
    p_{K_e}(k)=\int_\mathcal H\frac{e^{-H}H^k}{k!}f_H(H)\,\mathrm dH=\frac{(-1)^k}{k!}\partial_t^kM_H(-t)|_{t=1},
\end{equation}
where $M_H(t)=\mathsf Ee^{tH}$ denotes the moment generating function of the quanta exposure random variable $H$.


\section{Experimental Method} 

The experimental data was collected using a developmental DSERN capable camera from Gigajot Technology Inc. The specific camera chosen is the GJ00111, which consists of a monochrome one megapixel CMOS QIS with $1.1\,\mu\mathrm m$ pitch pixels. It was operated at its full bit-depth of 14-bits using four Correlated Multi-Sample (CMS) cycles to minimize read noise.

For this experiment, the PCH-EM algorithm was used to  estimate per-pixel dark current. This is accomplished through operating the camera with a lens cap and using a long integration time of $t_\mathrm{int}=4.87\,\mathrm s$. The long integration time ensures each pixel was given ample opportunity to produce thermally generated free-electrons. A total of $17,750$ frames over a $512\times 512\,\mathrm{px}$ region of interest were captured continuously under the dark environment.


\section{Results}

The PCH-EM algorithm code, available on the Mathworks File Exchange \cite{PCHEM_code}, was applied on a per pixel basis to the experimental dataset. To expedite calculations, parallel methods (memory limitations permitting) can be used as the per-pixel estimates can be found independently. An additional speed improvement is also possible by implementing PCH-EM through histograms (number of occurrences for each unique DN observed). Using the histogram is especially beneficial when there are relatively few unique DN values in a sample compared to the number of frames reported. Finally, an additional improvement can also be achieved by vectorizing the code, running the same sequence of calculations on multiple pixels simultaneously with MATLAB's optimized methods. The final time for the analysis was slightly under an hour running 12 cores on the machine used. Eventually, the release of the optimized histogram implementation of PCH-EM and other improvements to the algorithm in future updates will be provided on the Mathworks file exchange. 

The experiments were conducted with no external illumination with the intention of characterizing the dark current. The dark current $(i_d)$ given in units of $(e\text{-}/\mathrm{px}/\mathrm s)$ is found from the relation $i_d=H/t_\mathrm{int}$. Additionally, the read plus quantization noise $(\sigma_{e\text{-}})$ given in units of $(e\text{-})$ is found by multiplying the gain by the square root of $\sigma^2$, i.e.~$\sigma_{e\text{-}}=\sqrt{\sigma^2}\times g$.


\subsection{Per-pixel Characterization}

\begin{figure}[htb]
\centering
\includegraphics[scale=0.9]{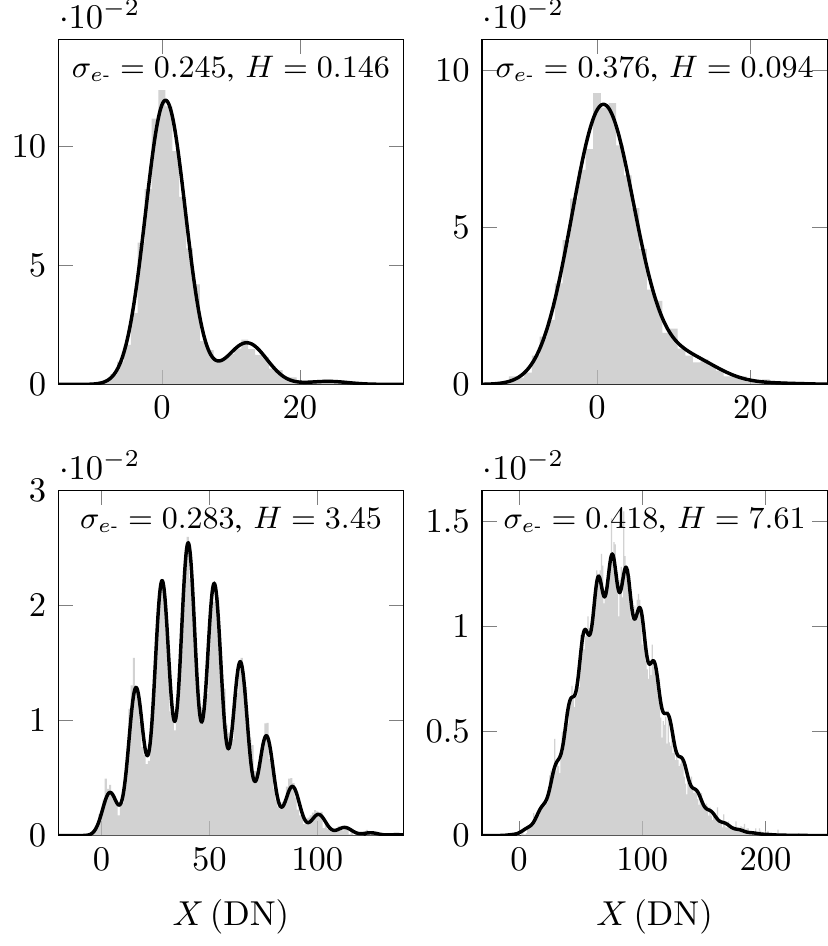}
\caption{Experimental per-pixel histograms for various read noise and quanta exposure. Histograms are fit with the PCD model using the PCH-EM algorithm.}
\label{fig:plot_per_pixel_fits}
\end{figure}

The PCH-EM algorithm provides estimates of the PCD model parameters $H$, $g$, $\mu$, and $\sigma^2$. Using these estimated parameters, the predicted probability density of the individual pixels can be computed. A comparison of the predicted density (solid black line) against the observed experimental histogram (gray bars) for four of the sensor's pixels are shown in Figure \ref{fig:plot_per_pixel_fits}.   

The four pixels shown were selected to demonstrate that the algorithm provides a good fit to the data at high or low quanta exposure as well as high or low read noise. Note that with low quanta exposure, there are very few peaks that may be used for estimating the conversion gain. However, even under such conditions, the probability density function calculated from the parameter estimates still accurately matches the observed data histogram. Also note that since these are dark frame measurements, the quanta exposure represents the expected number of free-electrons generated per-integration time via thermal contributions. Since this is proportional to the integration time, increasing the integration time will increase the observed quanta exposure and can further improve the estimates of the conversion gain if needed.


\subsection{PCD Parameter Maps}

Applied to the array, the PCD parameters for each pixel were estimated resulting in four two-dimensional arrays (maps) containing per-pixel estimates of $H$, $g$, $\mu$, and $\sigma^2$. Perhaps the most important for DSERN sensors is the distribution of read noise shown below in Figure \ref{fig:sigma_r_dist}. As can be seen, the vast majority of the pixels have an estimated read noise of less than $0.4\,e\text{-}$ with the median of the histogram occurring at $0.305\,e\text{-}$.
\begin{figure}[htb]
\centering
\includegraphics[scale=0.9]{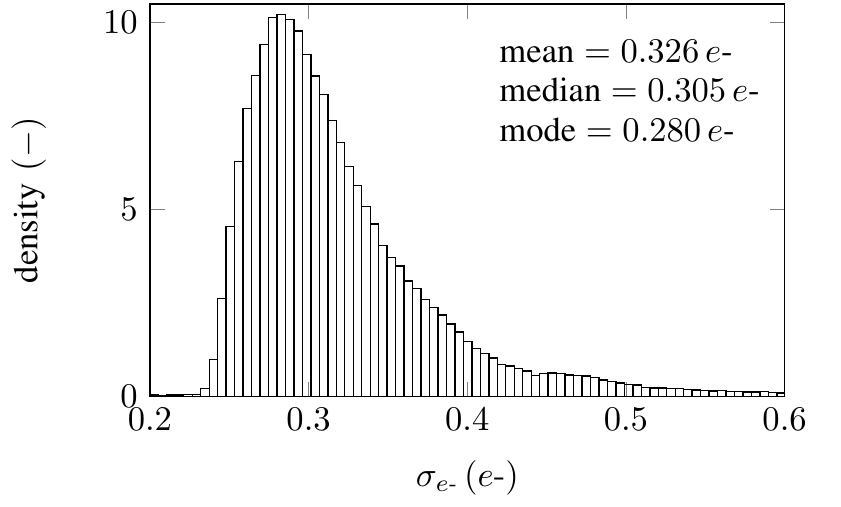}
\caption{Experimental histogram of read noise $(\sigma_{e\text{-}})$ estimates obtained from the PCH-EM algorithm.}
\label{fig:sigma_r_dist}
\end{figure}

The spatial context and distributions of other parameters are found in Figures \ref{fig:iD_map}-\ref{fig:sigmaR_map} (see Appendix \ref{sec:estimated_maps}). Structure (or lack thereof) observed in the parameter maps can be tied back to the architecture of the sensor and may potentially be useful in tuning the sensor parameters during development.


\subsection{Ensemble Distributions}
\label{subsec:experimental_ensemble_dist}

Applying the parameter estimates through (\ref{eq:full_ensemble_dist}), one can observe how the PCH-EM algorithm fits the sensor data on the array scale by predicting the EPCD of the sensor and comparing it to the ensemble histogram of the raw data. In order to estimate the EPCD, the unknown joint density of the PCD parameters $f_\theta$ must be determined. While this density is unknown, it may be approximated by binning the four parameter maps in a four-dimensional histogram. After normalization, this provides a discrete approximation for $f_\theta$. The approximate EPCD is then found by evaluating (\ref{eq:full_ensemble_dist}), replacing integrals with sums, for an appropriate range of $x$-values.

\begin{figure}[htb]
\centering
\includegraphics[scale=0.9]{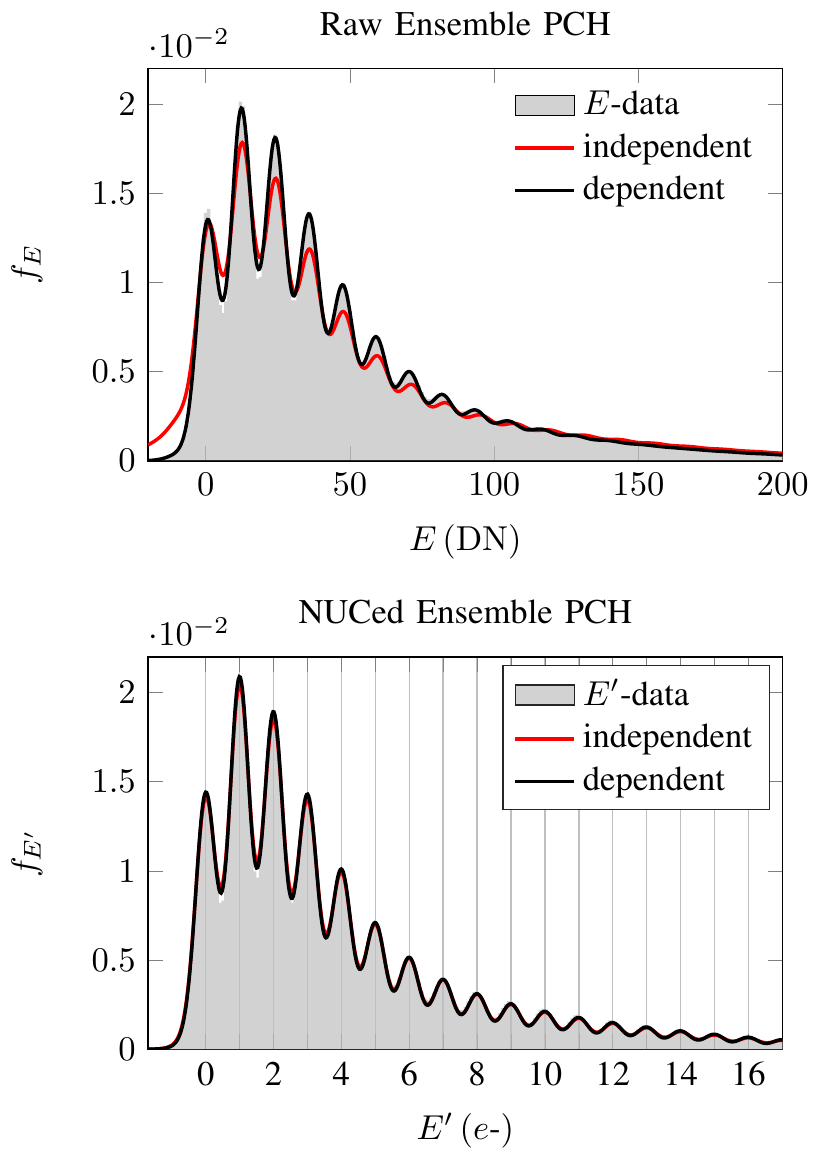}
\caption{Experimental ensemble histogram compared to estimated EPCD for the case of mutually independent and dependent parameters (top), and corresponding NUCed ensemble histogram with predicted NUCed EPCD (bottom).}
\label{fig:ensemble_plot}
\end{figure}

Figure \ref{fig:ensemble_plot} (top) shows the ensemble histogram made from $250$ frames of the raw experimental data compared to the estimated EPCD using the parameter maps. For comparison, two EPCD's were estimated under the assumption of mutually independent parameters ($f_\theta$ approximated by the product of four individual histograms) and dependent parameters ($f_\theta$ approximated by a single four-dimensional histogram), respectively. One can see that the EPCD under the assumption of dependent parameters provides an excellent fit to the raw data; thus providing experimental confirmation of the PCD model and PCH-EM algorithm. The fact that the estimated EPCD for dependent parameters ($\operatorname{RSME}=1.9\times 10^{-4}$) provides a better fit compared to the case of independent parameters ($\operatorname{RSME}=8.3\times 10^{-4}$) makes sense since, for example, the expression for the variance contains $g$; therefore, it is expected for $\sigma^2$ and $g$ to be dependent. In the ensemble histogram, it can be observed that the peaks become less distinct as signal increases which is usually indicative of conversion gain nonuniformity (see Appendix \ref{sec:ensemble_peak_resolution}). 

Figure \ref{fig:ensemble_plot} (bottom) shows the NUCed ensemble histogram found by subtracting per-pixel estimates of $\mu$ from each frame and then multiplying the bias corrected frames by the per-pixel estimates of $g$. This effectively removes the effects of gain and offset nonuniformity from the raw data. Notice that the peaks are now more clearly resolved and located at nonnegative integers showing that this two-point NUC restores the electron counting capabilities of the sensor. Using the same approach as before, the NUCed EPCD can be found by approximating the joint density $f_{\theta^\prime}$ from the quanta exposure and read noise maps under the assumption of dependent and independent parameters, and then approximating the double integral in (\ref{eq:partial_ensemble_dist}) by sums. As seen in the bottom of Figure \ref{fig:ensemble_plot}, both ensemble predictions under the assumption of dependent parameters ($\operatorname{RSME}=2.3\times 10^{-4}$) and independent parameters ($\operatorname{RSME}=2.7\times 10^{-4}$) fit the NUCed data quite well with a slight advantage given to the case of dependent parameters. This indicates that the quanta exposure (dark current) is nearly independent of the read noise (when in units of electrons), which is to be expected (see discussion at the end of Appendix \ref{sec:ensemble_peak_resolution}).

\begin{figure}[htb]
\centering
\includegraphics[scale=0.9]{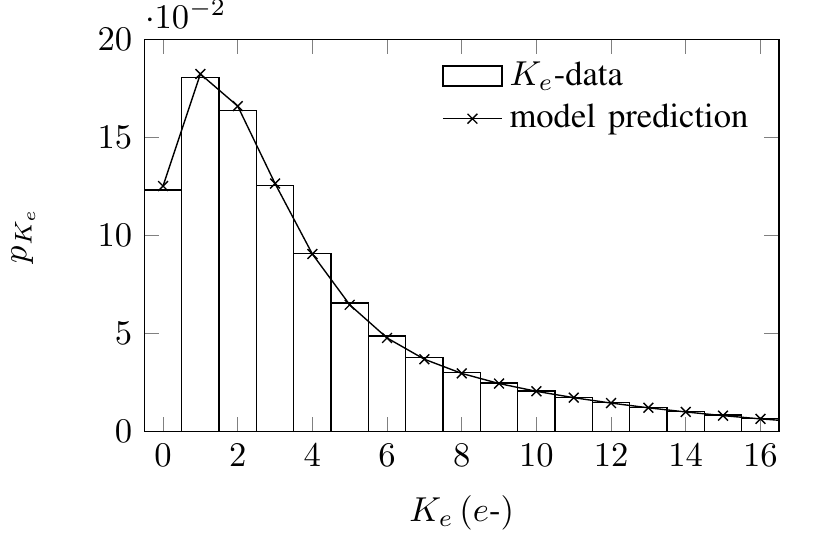}
\caption{Ensemble histogram of predicted electron numbers compared to the theoretical probability mass (\ref{eq:K_ensemble_pmf}) estimated via the quanta exposure map.
}
\label{fig:poisson_demarg}
\end{figure}

Through Figures \ref{fig:plot_per_pixel_fits} and \ref{fig:ensemble_plot}, the PCD and PCH-EM algorithm have been shown to be effective in modeling DSERN sensor data and providing estimates of the model parameters, respectively. What remains to be demonstrated is if the electron number prediction formula in (\ref{eq:demarg_K_eqn}) can effectively recover the electron numbers for each observation. Using (\ref{eq:demarg_K_eqn}), the predicted electron number for each pixel of the $250$-frame stack of raw experimental data was computed. This process resulted in an array of nonnegative integers, the same size as the image stack, containing the all predictions. A histogram of the predictions is given in Figure \ref{fig:poisson_demarg}. While it cannot be known if these predictions agree with the actual electron numbers associated with each observation, the distribution of the predictions can be compared to what would be expected according to the ensemble electron number probability mass in (\ref{eq:K_ensemble_pmf}). To predict this ensemble distribution, the unknown quanta exposure density $f_H$ was approximated by binning the quanta exposure map and then replacing the integral in (\ref{eq:K_ensemble_pmf}) by a finite sum. Figure \ref{fig:poisson_demarg} compares the ensemble histogram of the electron number predictions against the predicted probability mass according to the model. Recalling the discussion in Section \ref{subsec:The_PCH-EM_algorithm}, the data presented in Figure \ref{fig:poisson_demarg} can be viewed as an optimal quantization of the NUCed EPCD in Figure \ref{fig:ensemble_plot} (bottom). The quality of fit between the data and predicted probability mass demonstrates, at the very least, that the predicted electron numbers agree with the actual electron numbers in terms of distribution.


\section{Discussion and Future Work}

In this paper, the PCH-EM algorithm proposed in \cite{hendrickson_2023} was successfully demonstrated to accurately estimate quanta exposure, conversion gain, bias, and read noise of DSERN pixels in an automated fashion. Combining the assumed model with the corresponding estimated parameters accurately recreates the raw sensor data histograms, both on a per-pixel level as well as at the ensemble (array) level. The ensemble prediction required accounting for the correlation of the four model parameters. Additionally, it was shown how a two-point non-uniformity correction may be determined and applied to the ensemble, which improves the resolution of individual electron peaks and restores electron counting of the device. Lastly, the ability of PCH-EM to denoise raw sensor measurements and recover the hidden electron signal was demonstrated.

This PCH-EM algorithm is a powerful tool for investigating and tuning the performance of DSERN sensors, as it can be applied automatically over a large span of parameters. Through the use of the estimated parameter maps, PCH-EM not only is useful for sensor characterization but also may find application during the advanced development of the sensors themselves. Also, together with the Monte Carlo methods provided in \cite{PCHEM_code}, an experimentalist can investigate the number of frames required to achieve a desired uncertainty.

Future work will include expanding upon the current method to combine multiple illumination level measurements in a multi-sample version of PCH-EM, exploring techniques for accounting for non-linear responses, and releasing optimized code on the Mathworks File Exchange. Additionally, implementing various techniques for estimating the sample Fisher information will be pursued \cite{Louis_1982,Meng_1991,Oakes_1999,Meng_2017}. The ability to estimate the Fisher information would allow the PCH-EM algorithm to not only provide the parameter estimates but also their uncertainties. Ultimately, a generalized characterization method should work across the full dynamic range of the sensor and full parameter space of photon counting sensors.


\section{Acknowledgments}
\noindent The authors would like thank Nico Schl\"{o}mer for his {\ttfamily{matlab2tikz}} function, which was used to create the figures throughout this work \cite{schlomer_2021}.


\appendices

\section{Moments of the Ensemble Distributions}
\label{sec:ensemble_dist_moments}

Moments of the EPCD are found by noting that $E|\theta\sim\operatorname{PCD}(H,g,\mu,\sigma^2)$. Using the law of total expectation the first moment is
\begin{equation}
    \mathsf E(E)=\mathsf E(\mathsf E(E|\theta))=\mathsf E(\mu)+\mathsf E(H/g).
\end{equation}
Likewise, by the law of total variance
\begin{equation}
\begin{aligned}
    \mathsf{Var}(E)
    &=\mathsf{Var}(\mathsf E(E|\theta))+\mathsf E(\mathsf{Var}(E|\theta))\\
    &=\mathsf{Var}(\mu+H/g)+\mathsf E(\sigma^2)+\mathsf E(H/g^2).
\end{aligned}
\end{equation}
Expanding the first variance term further then gives the final result of
\begin{multline}
    \mathsf{Var}(E)
    =\mathsf{Var}(\mu)+\mathsf{Var}(H/g)+\mathsf E(\sigma^2)+\mathsf E(H/g^2)\\
    +2(\mathsf E(\mu H/g)-\mathsf E(\mu)\mathsf E(H/g)).
\end{multline}
The analogous moments of the NUCed EPCD come from these expressions upon setting $\mu=0$ and $g=1$ as constants. This gives $\mathsf E(E^\prime)=\mathsf E(H)$ and $\mathsf{Var}(E^\prime)=\mathsf{Var}(H)+\mathsf E(H)+\mathsf E(\sigma_{e\text{-}}^2)$.


\section{Dependence of Ensemble Peak Resolution on Parameter Nonuniformity}
\label{sec:ensemble_peak_resolution}

Here, the loss of peak resolution in the EPCD at higher signal levels and the dependence of this phenomenon on parameter nonuniformity is investigated. To do this, it is important to first understand why this behavior is not observed in the single pixel model.

Recall the distribution for a single pixel is given by the PCD
\begin{equation}
    f_X(x|\theta)=\sum_{k=0}^\infty\mathsf P(K=k)\underbrace{\phi(x;\mu+k/g,\sigma^2)}_{f_{X|K}(x|k)},
\end{equation}
which is comprised of an infinite mixture of Gaussian \emph{components} given by the probability density $f_{X|K}$. The individual components are thus isolated by considering the distribution of the random variable $X|K=k\sim\mathcal N(\mu+k/g,\sigma^2)$. Computing the variance of this conditioned variable gives
\begin{equation}
    \mathsf{Var}(X|K=k)=\sigma^2,
\end{equation}
which is independent of the electron number $k$. This means that the widths of each component making up the PCD are the same; thus, as signal $(k)$ increases, the resolution of individual peaks remains constant.

Repeating this calculation for the ensemble variable $E$, while assuming the appropriate regularity conditions to allow interchanging series and integration, the EPCD in (\ref{eq:full_ensemble_dist}) can be written in the form
\begin{equation}
    \label{eq:EPCD_alt_form}
    f_E(x)=\sum_{k=0}^\infty\mathsf P(K_e=k)\underbrace{\frac{\mathsf E_\theta(e^{-H}H^k\phi(x;\mu+k/g,\sigma^2))}{\mathsf E_\theta(e^{-H}H^k)}}_{f_{E|K_e}(x|k)},
\end{equation}
which is comprised of an infinite mixture of non-Gaussian components given by the probability density $f_{E|K_e}$. The variance of the conditioned variable $E|K_e=k\sim f_{E|K_e}$ is then given by
\begin{equation}
    \mathsf{Var}(E|K_e=k)=\mathsf E(E^2|K_e=k)-(\mathsf E(E|K_e=k))^2,
\end{equation}
where
\begin{equation}
    \mathsf E(E^2|K_e=k)=\frac{\mathsf E_\theta(e^{-H}H^k(\sigma^2+(\mu+k/g)^2)}{\mathsf E_\theta(e^{-H}H^k)}
\end{equation}
and
\begin{equation}
    \mathsf E(E|K_e=k)=\frac{\mathsf E_\theta(e^{-H}H^k(\mu+k/g))}{\mathsf E_\theta(e^{-H}H^k)}.
\end{equation}
Upon inspection, $\mathsf{Var}(E|K_e=k)$ is dependent on $k$; thus the widths of the components comprising the EPCD vary with signal level leading to a loss of peak resolution at higher signals.

What is not clear is if the dependence of $\mathsf{Var}(E|K_e=k)$ on $k$ is linked to the nonuniformity of only a subset of the parameters. This can be investigated by considering what happens to $\mathsf{Var}(E|K_e=k)$ when holding none, one, two, three, or all four parameters constant. This results in sixteen cases. Evaluating all sixteen cases, it can be shown that holding $(H,g)$, $(H,g,\mu)$, $(H,g,\sigma^2)$, $(g,\mu,\sigma^2)$, or $(H,g,\mu,\sigma^2)$ constant removes the dependence on $k$. Since the $(H,g)$ case implies the $(H,g,\mu)$ and $(H,g,\sigma^2)$ cases, and the $(H,g,\mu,\sigma^2)$ case results in the original PCD, there are only two ways for $\mathsf{Var}(E|K_e=k)$ to be independent of $k$ under dependent parameters: when $(H,g)$ is constant or $(g,\mu,\sigma^2)$ is constant. Thus holding certain subsets of the parameters constant does remove the dependence of $\mathsf{Var}(E|K_e=k)$ on $k$ resulting in constant peak resolution. It is also worth noting that $g$ appears in all of these cases showing that if conversion gain nonuniformity exists, then the EPCD component width must depend on $k$; causing peak resolution to decrease at higher signal levels. It is interesting that holding only $g$ constant does not remove the dependence on $k$; however, note that if $H$ is independent of $(g,\mu,\sigma^2)$ and then $g$ is held constant $\mathsf{Var}(E|K_e=k)=\mathsf E(\sigma^2)+\mathsf{Var}(\mu)$. This shows that the loss of resolution in the EPCD peaks can be solely contributed to conversion gain nonuniformity of $H$ is independent of $(g,\mu,\sigma^2)$. With so many combinations to consider, a study of the statistical dependence of the individual parameters in actual sensor systems may help guide further analysis.

The component width of the NUCed EPCD can also be found as a special case of $\mathsf{Var}(E|K_e)$ for $\mu=0$ and $g=1$ constant. This special condition leads to
\begin{equation}
    \mathsf{Var}(E^\prime|K_e=k)=\frac{\mathsf E_\theta(e^{-H}H^k\sigma_{e\text{-}}^2)}{\mathsf E_\theta(e^{-H}H^k)}\overset{H\perp\sigma_{e\text{-}}^2}{=}\mathsf E_\theta(\sigma_{e\text{-}}^2),
\end{equation}
where the last equality holds when $H$ is independent of $\sigma_{e\text{-}}^2$. This explains why the component width of the NUCed EPCD in Figure \ref{fig:ensemble_plot} appears to be constant.


\section{Estimated Parameter Maps}
\label{sec:estimated_maps}

One of the challenges when displaying the estimated parameter maps is the presence of outliers, which given a limited dynamic range of the display means one typically has to clip the map values. To provide a visually aesthetic way to display the maps, a nonlinear transformation of the form
\begin{equation}
    x_{ij}^\prime=F^{-1}_{\beta(2,2)}(\tilde F(x_{ij}))
\end{equation}
was applied to the map elements. Here, $F^{-1}_{\beta(2,2)}$ is the $\operatorname{Beta}(2,2)$ quantile function, $\tilde F$ is the empirical cumulative distribution function of the map, and $x_{ij}$ is the $ij$th element of the map. This transformation takes the original histogram of the map and shapes it into that of a $\operatorname{Beta}(2,2)$ distribution; however, because this transformation is monotone, any structures in the original map are carried over to the final transformation, all while suppressing the appearance of outliers so that the structure is clearly observed.

\begin{figure}[p]
\centering
\includegraphics[scale=0.7]{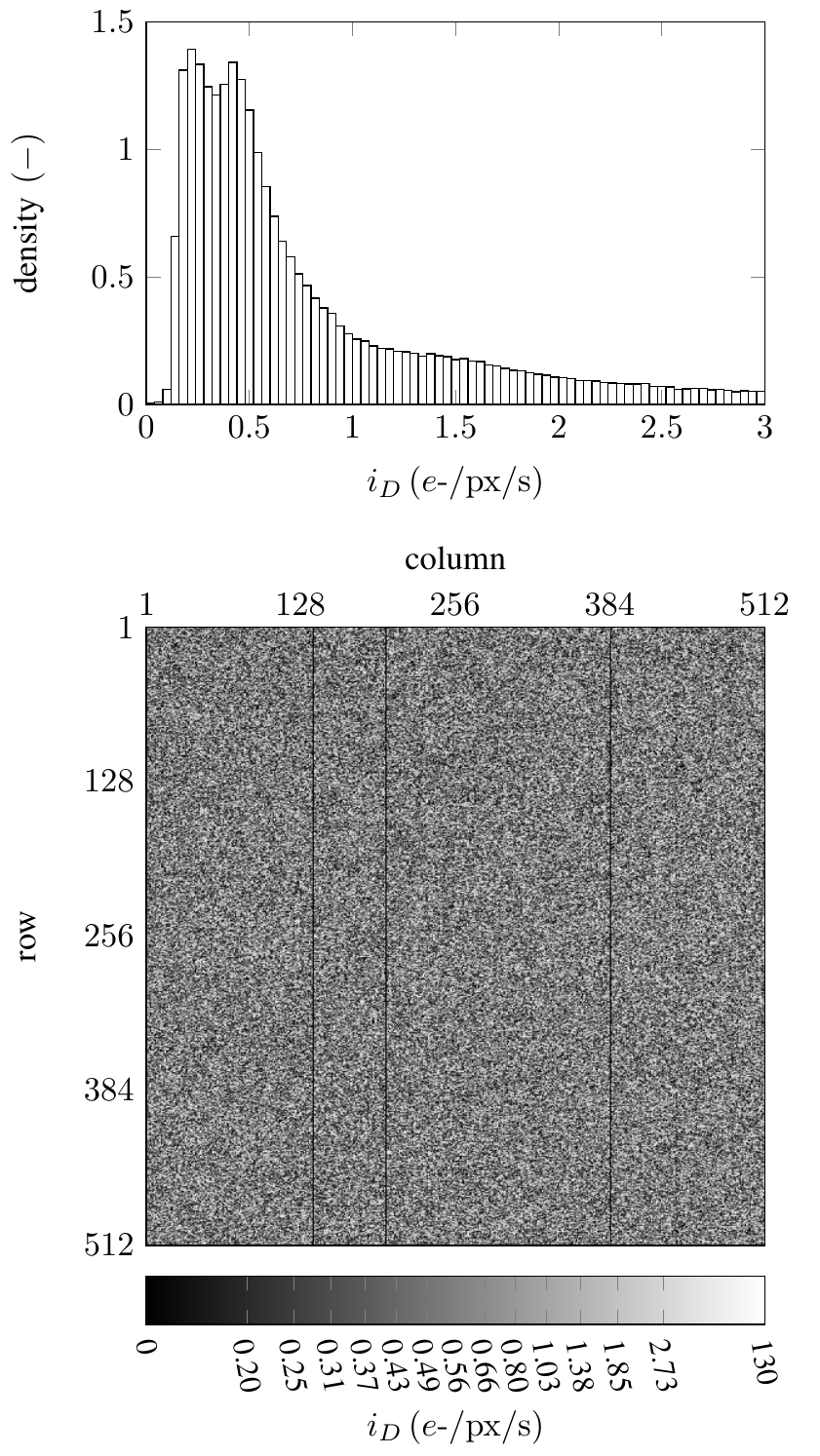}
\caption{Experimental histogram of dark current $(i_d)$ estimates obtained from the PCH-EM algorithm (top) with corresponding map (bottom).}
\label{fig:iD_map}
\end{figure}

\begin{figure}[p]
\centering
\includegraphics[scale=0.7]{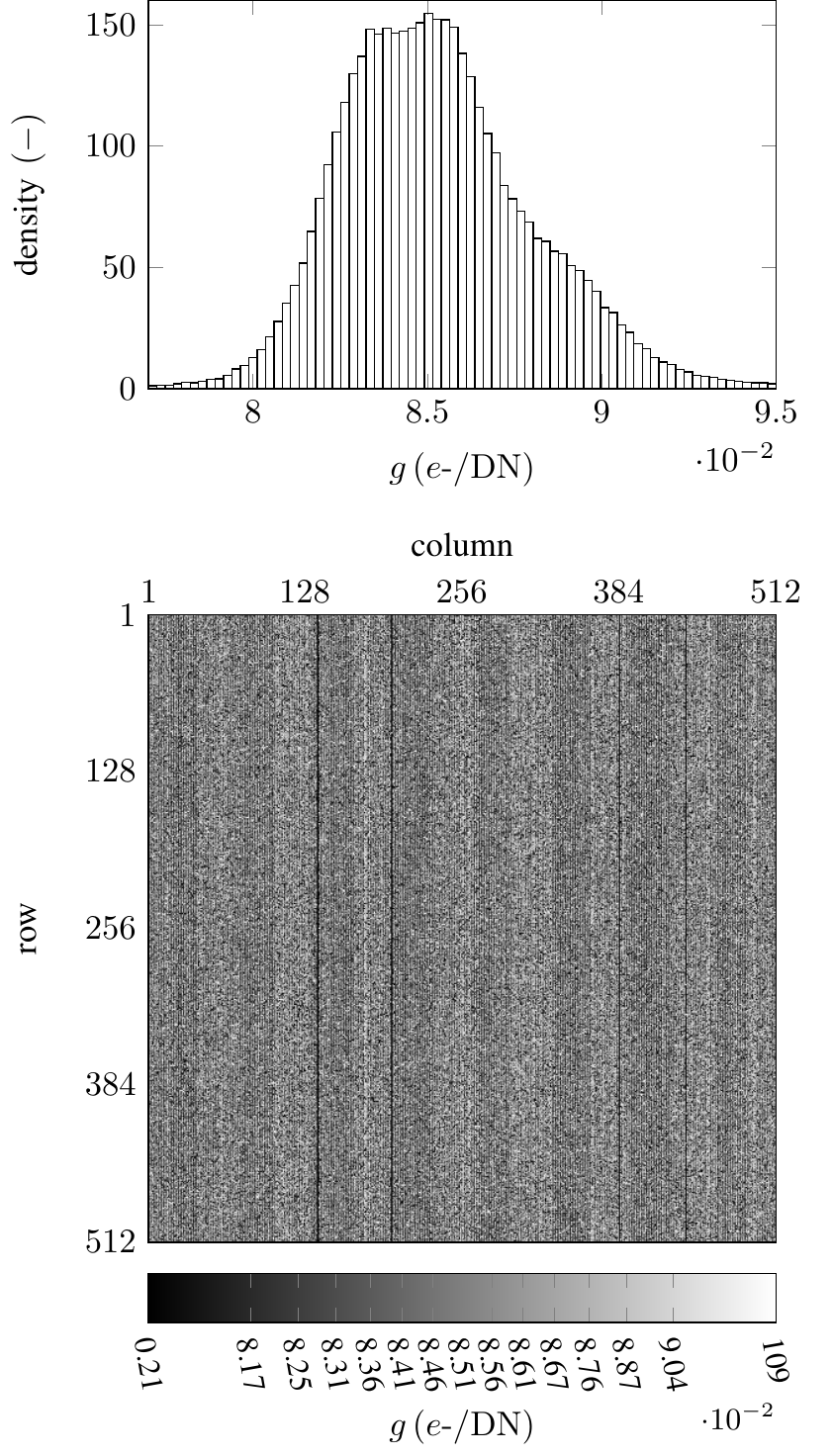}
\caption{Experimental histogram of conversion gain $(g)$ estimates obtained from the PCH-EM algorithm (top) with corresponding map (bottom).}
\label{fig:g_map}
\end{figure}

\begin{figure}[p]
\centering
\includegraphics[scale=0.7]{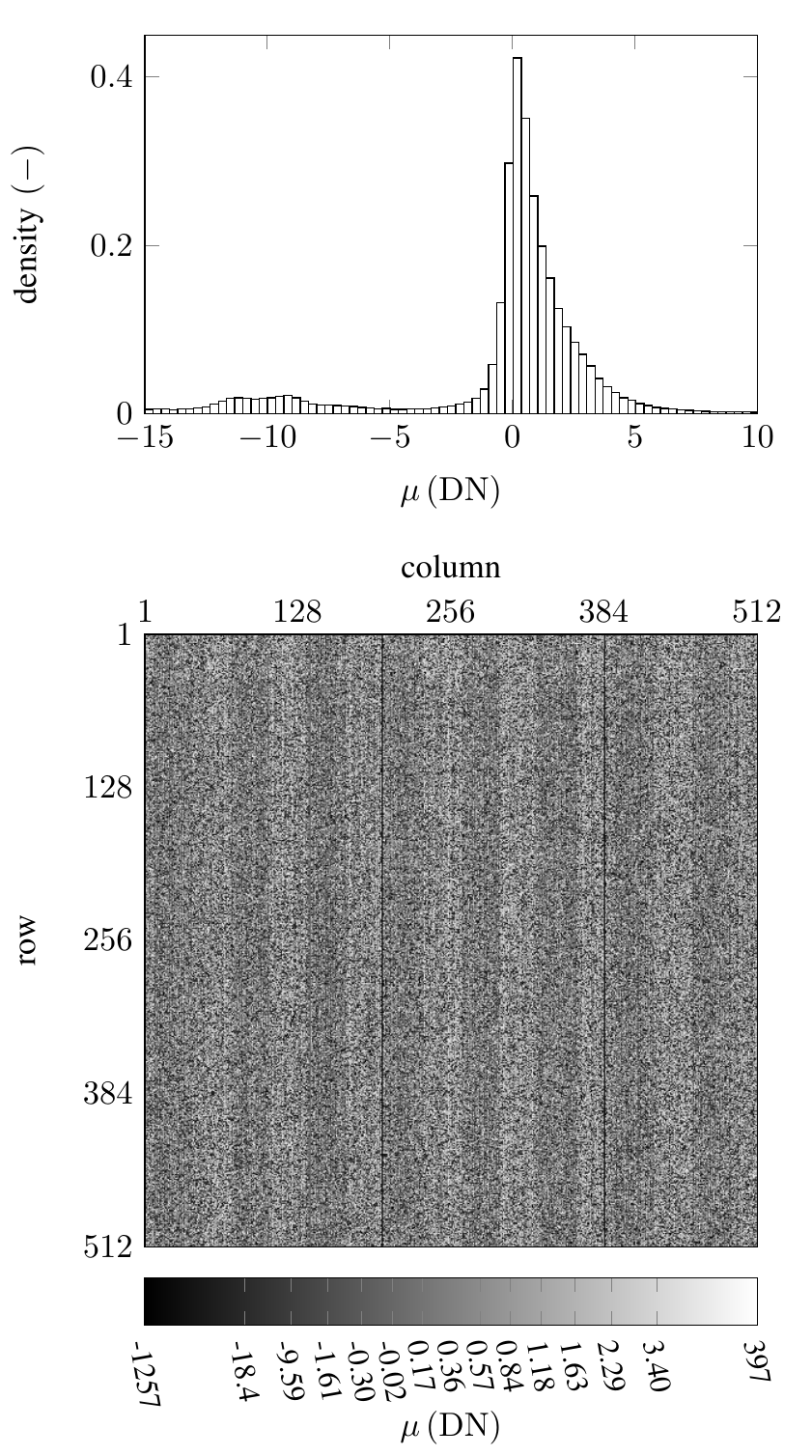}
\caption{Experimental histogram of bias $(\mu)$ estimates obtained from the PCH-EM algorithm (top) with corresponding map (bottom).}
\label{fig:mu_map}
\end{figure}

\begin{figure}[p]
\centering
\includegraphics[scale=0.7]{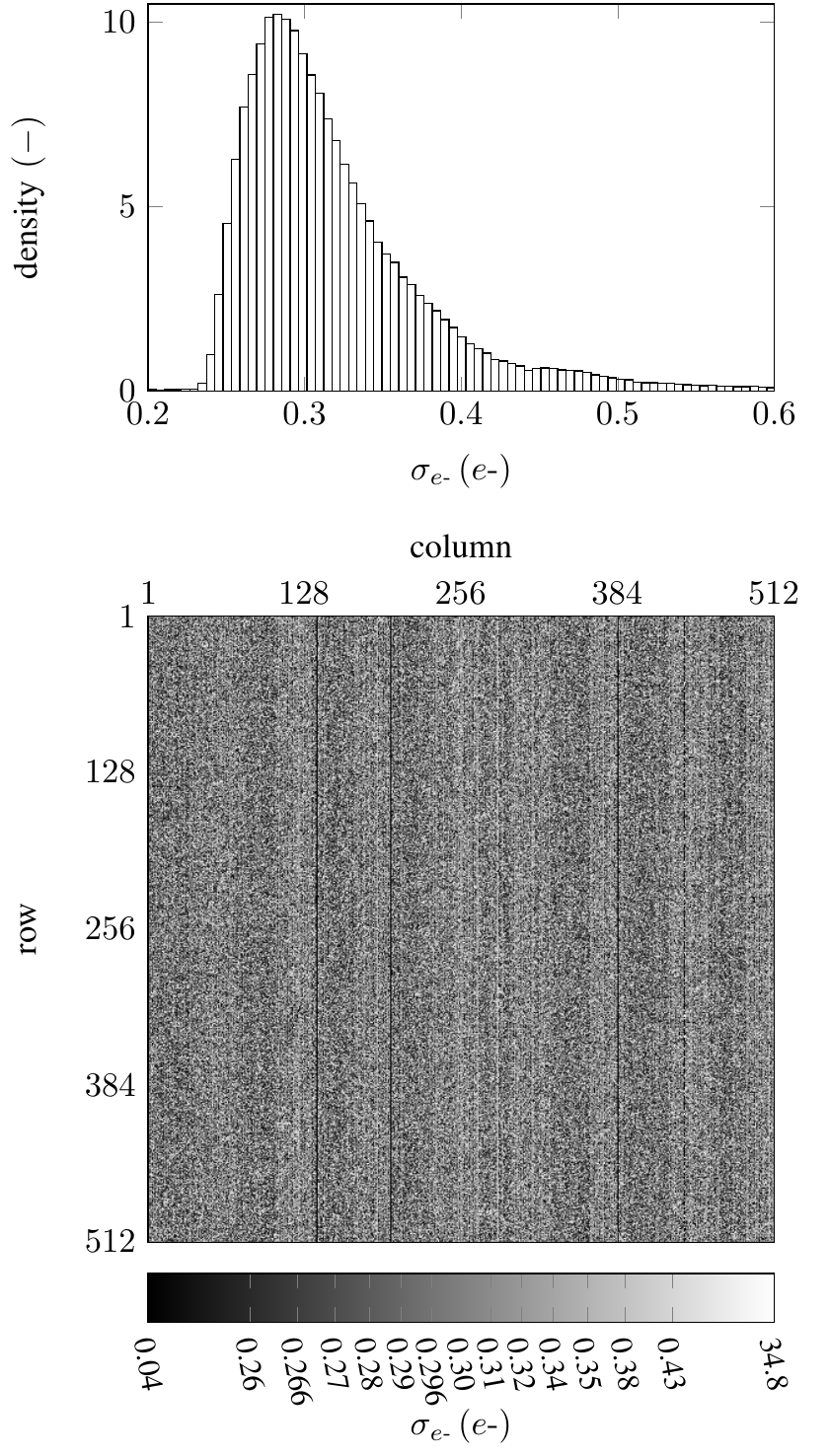}
\caption{Experimental histogram of read noise $(\sigma_{e\text{-}})$ estimates obtained from the PCH-EM algorithm (top) with corresponding map (bottom).}
\label{fig:sigmaR_map}
\end{figure}


\newpage

\begin{IEEEbiography}[{\includegraphics[width=1in,height=1.25in,clip,keepaspectratio]{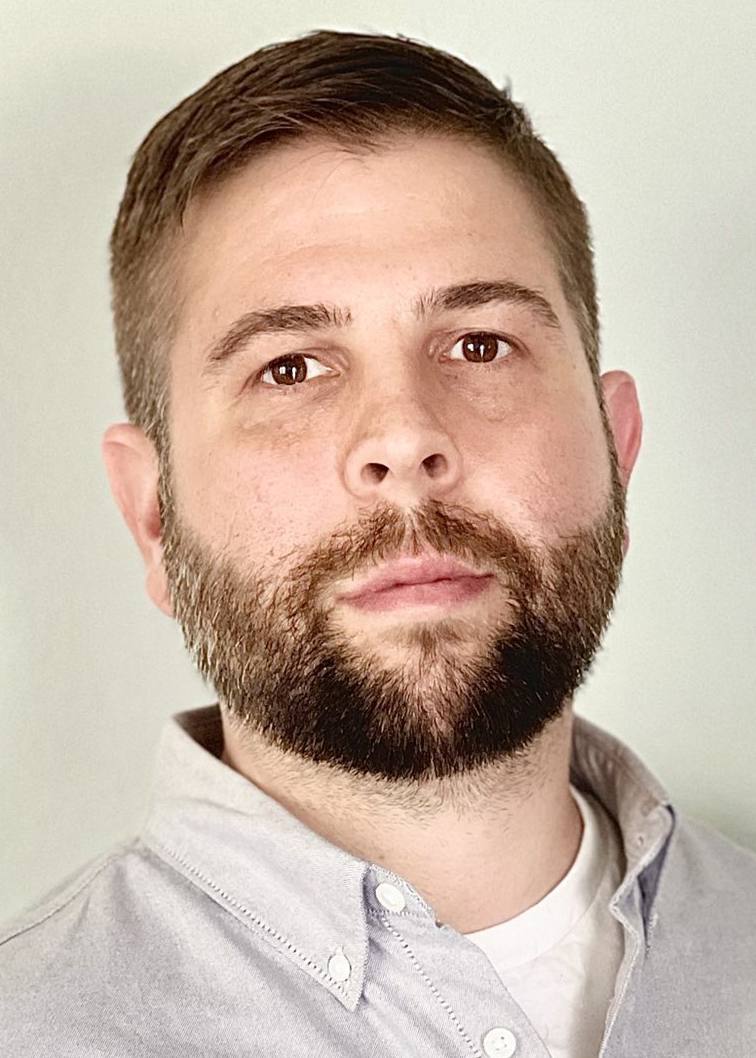}}]{Aaron J.~Hendrickson} received the B.S.~degree in Imaging and Photographic Technology from the Rochester Institute of Technology, Rochester, NY, USA, in 2011, and the M.S.~degree in Applied and Computational Mathematics from Johns Hopkins University, Baltimore, MD, USA, in 2020. He is currently working for the U.S.~Department of Defense. His research focus is in developing theoretical foundations for image sensor characterization methods. 
\end{IEEEbiography}

\begin{IEEEbiography}[{\includegraphics[width=1in,height=1.25in,clip,keepaspectratio]{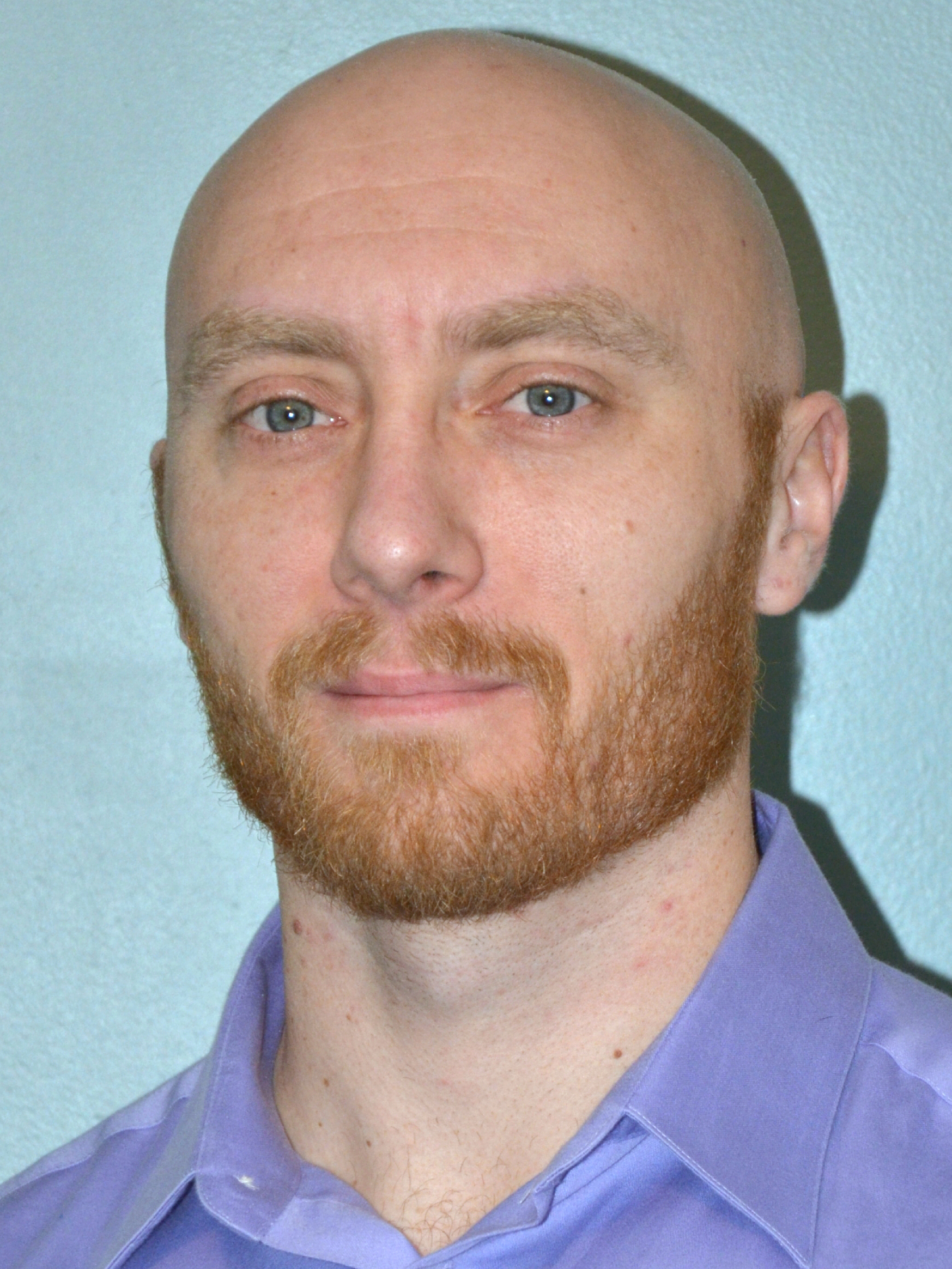}}]{David P.~Haefner} received his B.S.~in Physics from ETSU in 2004, a Ph.D.~in Optics from the UCF’s CREOL in 2010, a M.S.~in Electrical Engineering, and a M.S.~in Mechanical Engineering from CUA in 2014 and 2015, respectively. Since 2010 he has worked at the U.S. Army Combat Capabilities Development Command (DEVCOM) C5ISR Center. His current research spans electro-optic imaging system measurement for performance predictions and new measurement development.
\end{IEEEbiography}

\begin{IEEEbiography}[{\includegraphics[width=1in,height=1.25in,clip,keepaspectratio]{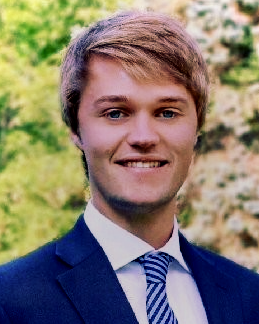}}]{Nicholas R.~Shade} received his A.B and B.E in electrical engineering from the Thayer School of Engineering, Dartmouth College, in 2020. He is currently a Ph.D.~candidate at Thayer working in Eric Fossum's advanced camera technology lab.
\end{IEEEbiography}

\begin{IEEEbiography}[{\includegraphics[width=1in,height=1.25in,clip,keepaspectratio]{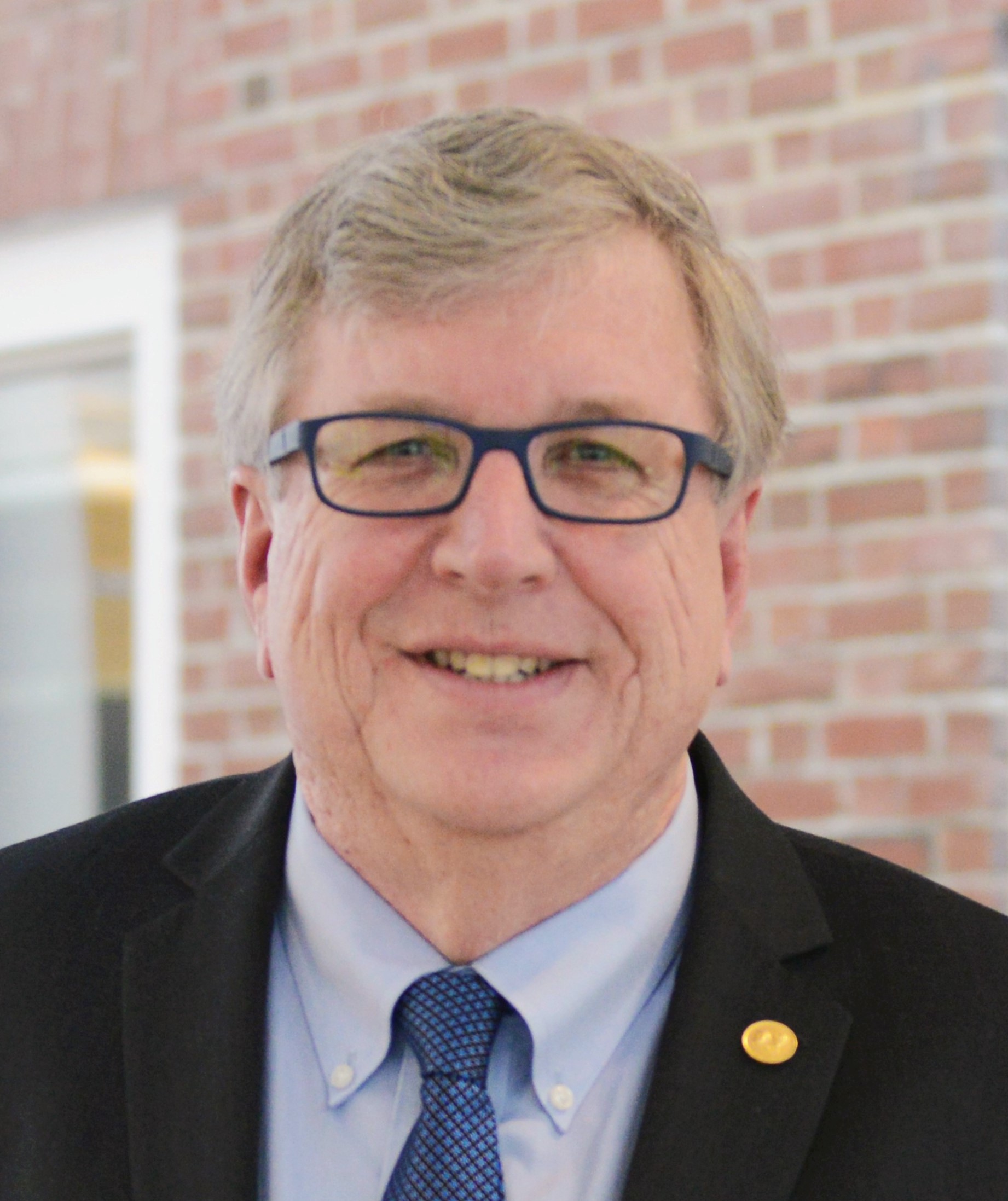}}]{Eric R.~Fossum} is the John H.~Krehbiel Sr.~Professor for Emerging Technologies with the Thayer School of Engineering, Dartmouth College. He is the primary inventor of the CMOS image sensor used in billions of smartphones and other applications and is currently exploring the quanta image sensor. He was awarded the IEEE Andrew Grove Award, the Queen Elizabeth Prize for Engineering in 2017, the OSA/IS\&T Land Medal in 2020, and the Technical Emmy Award in 2021, among other honors. He was the Co-Founder and the First President of the International Image Sensor Society. He was inducted into the National Inventors Hall of Fame and is a Member of the National Academy of Engineering.
\end{IEEEbiography}


\bibliographystyle{IEEEtran}
\bibliography{sources}

\begin{thebibliography}{10}
\providecommand{\url}[1]{#1}
\csname url@samestyle\endcsname
\providecommand{\newblock}{\relax}
\providecommand{\bibinfo}[2]{#2}
\providecommand{\BIBentrySTDinterwordspacing}{\spaceskip=0pt\relax}
\providecommand{\BIBentryALTinterwordstretchfactor}{4}
\providecommand{\BIBentryALTinterwordspacing}{\spaceskip=\fontdimen2\font plus
\BIBentryALTinterwordstretchfactor\fontdimen3\font minus
  \fontdimen4\font\relax}
\providecommand{\BIBforeignlanguage}[2]{{%
\expandafter\ifx\csname l@#1\endcsname\relax
\typeout{** WARNING: IEEEtran.bst: No hyphenation pattern has been}%
\typeout{** loaded for the language `#1'. Using the pattern for}%
\typeout{** the default language instead.}%
\else
\language=\csname l@#1\endcsname
\fi
#2}}
\providecommand{\BIBdecl}{\relax}
\BIBdecl

\bibitem{beecken_96}
B.~P. Beecken and E.~R. Fossum, ``Determination of the conversion gain and the
  accuracy of its measurement for detector elements and arrays,'' \emph{Appl.
  Opt.}, vol.~35, no.~19, pp. 3471--3477, Jul 1996.

\bibitem{janesick_2007}
J.~R. Janesick, \emph{Photon {T}ransfer: $DN\to\lambda$}.\hskip 1em plus 0.5em
  minus 0.4em\relax SPIE, 2007.

\bibitem{hendrickson_22}
A.~Hendrickson, D.~P. Haefner, and B.~L. Preece, ``On the optimal measurement
  of conversion gain in the presence of dark noise,'' \emph{J. Opt. Soc. Am.
  A}, vol.~39, no.~12, pp. 2169--2185, Dec 2022.

\bibitem{starkey_2016}
D.~A. Starkey and E.~R. Fossum, ``Determining conversion gain and read noise
  using a photon-counting histogram method for deep sub-electron read noise
  image sensors,'' \emph{IEEE Journal of the Electron Devices Society}, vol.~4,
  no.~3, pp. 129--135, May 2016.

\bibitem{Nakamoto_2022}
K.~Nakamoto and H.~Hotaka, ``Efficient and accurate conversion-gain estimation
  of a photon-counting image sensor based on the maximum likelihood
  estimation,'' \emph{Opt. Express}, vol.~30, no.~21, pp. 37\,493--37\,506, Oct
  2022.

\bibitem{hendrickson_2023}
A.~Hendrickson and D.~P. Haefner, ``Photon counting histogram expectation
  maximization algorithm for characterization of deep sub-electron read noise
  sensors,'' \emph{Cornell University arXiv}, vol. 2302.00090, 2023.

\bibitem{fossum_2015_2}
J.~Ma, D.~Starkey, A.~Rao, K.~Odame, and E.~R. Fossum, ``Characterization of
  quanta image sensor pump-gate jots with deep sub-electron read noise,''
  \emph{IEEE Journal of the Electron Devices Society}, vol.~3, no.~6, pp.
  472--480, 2015.

\bibitem{fossum_2015}
J.~Ma and E.~R. Fossum, ``Quanta image sensor jot with sub 0.3e- r.m.s. read
  noise and photon counting capability,'' \emph{IEEE Electron Device Letters},
  vol.~36, no.~9, pp. 926--928, 2015.

\bibitem{Dutton_2016}
N.~A.~W. Dutton, I.~Gyongy, L.~Parmesan, and R.~K. Henderson, ``Single photon
  counting performance and noise analysis of {CMOS} {SPAD}-based image
  sensors,'' \emph{Sensors}, vol.~16, no.~7, 2016.

\bibitem{dempster_1977}
A.~P. Dempster, N.~M. Laird, and D.~B. Rubin, ``{Maximum Likelihood from
  Incomplete Data Via the EM Algorithm},'' \emph{Journal of the Royal
  Statistical Society: Series B (Methodological)}, vol.~39, no.~1, pp. 1--22,
  1977.

\bibitem{fossum_2016_2}
E.~R. Fossum, ``Photon counting error rates in single-bit and multi-bit quanta
  image sensors,'' \emph{IEEE Journal of the Electron Devices Society}, vol.~4,
  no.~3, pp. 136--143, 2016.

\bibitem{PCHEM_code}
\BIBentryALTinterwordspacing
A.~Hendrickson and D.~P. Haefner, ``{One-Sample PCH-EM Algorithm},'' MATLAB
  Central File Exchange, 2022. [Online]. Available:
  \url{https://www.mathworks.com/matlabcentral/fileexchange/121343-one-sample-pch-em-algorithm}
\BIBentrySTDinterwordspacing

\bibitem{Louis_1982}
T.~A. Louis, ``Finding the observed information matrix when using the {EM}
  algorithm,'' \emph{Journal of the Royal Statistical Society: Series B
  (Methodological)}, vol.~44, no.~2, pp. 226--233, 1982.

\bibitem{Meng_1991}
X.~L. Meng and D.~B. Rubin, ``Using {EM} to obtain asymptotic
  variance-covariance matrices: The {SEM} algorithm,'' \emph{Journal of the
  American Statistical Association}, vol.~86, no. 416, pp. 899--909, 1991.

\bibitem{Oakes_1999}
D.~Oakes, ``Direct calculation of the information matrix via the {EM}
  algorithm,'' \emph{Journal of the Royal Statistical Society: Series B
  (Statistical Methodology)}, vol.~61, no.~2, pp. 479--482, 1999.

\bibitem{Meng_2017}
L.~Meng and J.~C. Spall, ``Efficient computation of the {F}isher information
  matrix in the {EM} algorithm,'' in \emph{2017 51st Annual Conference on
  Information Sciences and Systems (CISS)}, 2017, pp. 1--6.

\bibitem{schlomer_2021}
N.~Schl{\"{o}}mer, ``matlab2tikz: {A} script to convert {MATLAB}/{O}ctave into
  {T}ik{Z} figures for easy and consistent inclusion into {\LaTeX}.'' GitHub,
  {URL: https://github.com/matlab2tikz/matlab2tikz (retrieved May 8, 2021)}.

\end{thebibliography}

\end{document}